\documentclass[aps,twocolumn,showpacs,superscriptaddress,groupedaddress,nofootinbib]{revtex4-1} 
\usepackage{graphicx}
\usepackage[sort&compress]{natbib}
\usepackage{lipsum}
\usepackage{morefloats}
\usepackage[pdf]{pstricks}
\usepackage{subfigure}
\usepackage{amsmath}
\usepackage{amssymb}
\usepackage{amsfonts}
\usepackage{rotating}
\usepackage{cancel}
\usepackage{mathtools}
\usepackage{color}
\usepackage{bbm}
\usepackage{dsfont}
\usepackage{bbold}
\usepackage{multirow}
\usepackage{ulem}

\newcommand{\mev}{\textrm{ MeV}}

\begin{document}

\title{$D^0 D^0 \pi^+$ mass distribution in the production of the $T_{cc}$ exotic state}

\author{A. Feijoo}
\email{edfeijoo@ific.uv.es}
\affiliation{Departamento de F\'{\i}sica Te\'orica and IFIC,
Centro Mixto Universidad de Valencia-CSIC,
Institutos de Investigaci\'on de Paterna, Aptdo. 22085, 46071 Valencia, Spain}
\affiliation{Nuclear Physics Institute, 25068 Rez, Czech Republic}   

\author{W.~H.~Liang}
\email{liangwh@gxnu.edu.cn}
\affiliation{Department of Physics, Guangxi Normal University, Guilin 541004, China}
\affiliation{Guangxi Key Laboratory of Nuclear Physics and Technology, Guangxi Normal University, Guilin 541004, China}

\author{Eulogio Oset}
\email{Eulogio.Oset@ific.uv.es}
\affiliation{Departamento de F\'{\i}sica Te\'orica and IFIC,
Centro Mixto Universidad de Valencia-CSIC,
Institutos de Investigaci\'on de Paterna, Aptdo. 22085, 46071 Valencia, Spain}   
\affiliation{Department of Physics, Guangxi Normal University, Guilin 541004, China}

\begin{abstract}
We perform a unitary coupled channel study of the interaction of the  $D^{*+} D^0, D^{*0} D^+$ channels and find a state barely bound, 
very close to isospin $I=0$. We take the experimental mass as input and obtain the width of the state and the $D^0 D^0 \pi^+$ mass distribution. When the mass of the $T_{cc}$ state quoted in the experimental paper from raw data is used, the width obtained is of the order of the $80 \;{\rm keV}$, small compared to the value given in that work. Yet, when the mass obtained in an analysis of the data considering the experimental resolution is taken, the width obtained is about $43 \;{\rm keV}$ and both the width and the $D^0 D^0 \pi^+$ mass distribution are in remarkable agreement with the results obtained in that latter analysis.
\end{abstract}

\maketitle
\section{Introduction}
The recent discovery of the $T_{cc}$ state by the LHCb Collaboration \cite{exp1,exp2,exp3,LHCb:2021vvq} has added a new exotic hadron state
to an already long list of states discovered in the latest years that challenge the $q\bar q$ nature of the standard mesons or $qqq$ of the standard baryons.
The novelty with respect to many states containing hidden charm is that now there are two charm quarks open.
This finding follows the discovery of the $X_0(2866)$ and $X_1(2904)$
which have an open charm quark and a strange quark with manisfestly tetraquark structure \cite{x2866}.

On the theory side there have been quite a few works devoted to the study of tetraquarks with two heavy quarks \cite{1,2,3,4,5,9,10,11,12,13,14,15,150,16,17,18,19},
with quite a wide range of predictions going from about $250 \mev$ below the energy reported for the $T_{cc}$
to $250 \mev$ above in the case of two open charmed quarks.

The mass of the $T_{cc}$ state is remarkably close to the $D^{*+} D^0$ and $D^{*0} D^+$ thresholds, its value is given by \cite{exp1,exp2,exp3,LHCb:2021vvq}
\begin{equation}\label{eq:mexp}
m_{\rm exp} = 3875.09 \mev + \delta m_{\rm exp},
\end{equation}
where $3875.09 \mev$ is the threshold of the $D^{*+} D^0$ state and
\begin{equation}\label{eq:mexp2}
\delta m_{\rm exp}= -273\pm 61 \pm 5^{+11}_{-14}~{\rm keV}.
\end{equation}
The width reported for the $T_{cc}$ state is \cite{exp1,exp2,exp3,LHCb:2021vvq}
\begin{equation}\label{eq:mexp3}
\Gamma = 410 \pm 165 \pm 43^{+18}_{-38}~{\rm keV}.
\end{equation}
As we can see, the mass is very close to the $D^{*+} D^0$ threshold and the width is very small.

The results of  Ref.~\cite{LHCb:2021vvq} were accompanied by a theoretical analysis of the data by the LHCb Collaboration \cite{LHCb:2021auc}, using a unitary Breit-Wigner amplitude and taking into account the experimental resolution. The results obtained differ somewhat from those of Eqs.~\eqref{eq:mexp2} and \eqref{eq:mexp3} and the new values reported from the pole position of the state are 
\begin{equation}\label{eq:mpole}
\delta m_{\rm exp}= -360\pm 40^{+4}_{-0}~{\rm keV} ,
\end{equation}
\begin{equation}\label{eq:wpole}
\Gamma = 48 \pm 2^{+0}_{-14}~{\rm keV}.
\end{equation}

The closeness to the $D^* D$ threshold makes one think immediately about the possibility
that this state could be a molecular state of $D^* D$,
and in fact such structure was anticipated in Refs.~\cite{shilin,18,he}.
Independent of the structure of the $T_{cc}$ state,
the proximity of the $D^* D$ threshold makes unavoidable the explicit consideration of the $D^* D$ channels in its study,
as shown in the detailed study of threshold structures in Ref.~\cite{fengkun}.
The possible $D^* D$ bound state would have an analogous structure to the $D^*D^*$ molecular state already studied in Ref.~\cite{branz},
where such open charm molecular structures were reported for the first time.
It is interesting to mention that also in Ref.~\cite{branz} predictions were made for another exotic state of $D^* \bar K^*$ nature
that matches correctly the $X_0(2866)$ state reported in Ref.~\cite{x2866} (see update in Ref.~\cite{raquelnew}).

 The reaction of the theory community to the experimental finding has been fast.
 In Ref.~\cite{perfect} a reminder was given that in Ref.~\cite{shilin} a prediction for a molecular $D^* D$ state had been done matching perfectly the mass found in the experiment.
 In Ref.~\cite{shilingam} the width of the $T_{cc}$ state is studied with the $D^{*+} D^0$ and $D^{*0} D^+$ coupled channels and found small compared with the experimental one. \footnote{The couplings of $T_{cc}$ to the $D^{*+} D^0$ and $D^{*0} D^+$ channels obtained in Ref.~\cite{shilingam} (version v1 of the ArXiv) are under revision \cite{private}.}
 The same conclusion is obtained in Ref.~\cite{xiegeng} where a single channel $D^* D$ molecule is assumed.
 The QCD sum rules method also brings its contribution to the subject
 showing that such a state appears at a central value of $3868 \mev$ for the mass, 
 with the typical large uncertainties of the sum rules method,
 about $124 \mev$ in this case \cite{azizi}.

In the present work we report on how a molecular state of $D^{*+} D^0, D^{*0} D^+$ nature naturally emerges from the interaction of these two coupled channels,
and we make a study of the invariant mass distribution of $D^0 D^0 \pi^+$ in the production of this state, which is the mode where it has been observed in Refs.~\cite{exp1,exp2,exp3,LHCb:2021vvq}. We use as a source of interaction the exchange of vector mesons provided by the local hidden gauge approach \cite{hidden1,hidden2,hidden4,hideko}.
In the case of $VP$ (vector-pseudoscalar) interaction one can also have the exchange of pseudoscalar mesons, but comparatively to the vector exchange their contribution is very small \cite{diastoledo,aceti,nakamura} (see detailed calculations in Appendix~A of Ref~\cite{diastoledo}).
In any case the coupled channels unitary approach requires the use of the $G$ functions, the loop functions of the intermediate $D^* D$ states, which have to be regularized, and missing pieces of the interaction can be accommodated by means of an appropriate choice of the cut off or the subtraction constant, in the cut off or dimensional regularization methods, which are fine tuned to the precise value of the mass of the state.

\section{Formalism and Results}

\begin{figure*}
\centering
 \includegraphics[width=0.85\textwidth]{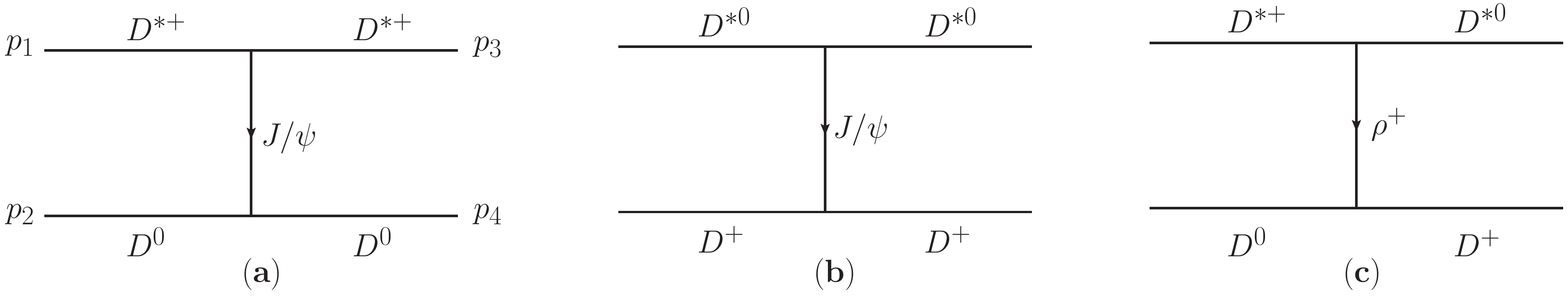}
 \caption{\small{Diagrams considered for the interaction $VP$. }}
\label{fig:FeynDiag}
 \end{figure*}

We use a unitary method with the coupled channels $D^{*+} D^0$ and $D^{*0} D^+$, paying attention to the exact masses and widths.
The interaction is obtained from the extended local hidden gauge Lagrangians \cite{hidden1,hidden2,hidden4,hideko}
and they correspond to the exchange of vector mesons in the diagrams of Fig.~\ref{fig:FeynDiag}.

The Lagrangians used are
\begin{eqnarray}
  {\cal{L}}_{VPP} &=& -ig \,\langle [P, \partial_\mu P] V^\mu\rangle, \nonumber \\
  {\cal{L}}_{VVV} &=& ig \,\langle (V^\nu \partial_\mu V_\nu-\partial_\mu V^\nu V_\nu) V^\mu\rangle, \\
  g &=& \frac{M_V}{2\,f}, ~~(M_V=800 \mev,~f=93 \mev).\nonumber
\end{eqnarray}
with $\langle~\rangle$ meaning the trace of the matrices in the SU(4) space,
where $P$ and $V$ stand for the pseudoscalars and vectors respectively,
and they correspond to the $q_i \bar q_j$ matrices written in terms of the corresponding mesons,
which can be found in Ref.~\cite{Ikeno}.
Since we are close to the $D^* D$ threshold we neglect the $\epsilon^0$ components of the vectors 
and work with the vector polarizations $\vec\epsilon, \vec\epsilon\,'$.
Calling $D^{*+} D^0, D^{*0} D^+$ the $1, 2$ channels, the interaction that we obtain is 
\begin{eqnarray}\label{eq:Vij}
  V_{ij} &=& C_{ij}\, g^2\, (p_1+p_3)\cdot (p_2+p_4) \, \vec\epsilon \cdot \vec\epsilon\,' \nonumber\\
   &\rightarrow &  C_{ij}\, g^2\, \frac{1}{2} [3s-(M^2+m^2+M'^2+m'^2)  \nonumber\\
   & & -\frac{1}{s}(M^2-m^2)(M'^2-m'^2)]\, \vec\epsilon \cdot \vec\epsilon\,',
\end{eqnarray}
where $M, m$ are the initial vector, pseudoscalar masses and $M', m'$ the corresponding final ones.
The second expression in Eq.~\eqref{eq:Vij} follows after projection in $s$-wave,
which is what we study.
The matrix $C_{ij}$ is given by
\begin{equation}\label{eq:Cij}
C_{ij} = \left(
           \begin{array}{cc}
             \frac{1}{M_{J/\psi}^2} & \frac{1}{m_\rho^2}   \\[0.1cm]
              \frac{1}{m_\rho^2}  &  \frac{1}{M_{J/\psi}^2}\\
           \end{array}
         \right).
\end{equation}

In diagrams Fig.~\ref{fig:FeynDiag}(a) and (b) one can exchange $\rho^0$ and $\omega$, but one can see that the product of the couplings for $\rho$ or $\omega$ exchange is the same, yet with opposite sign, and assuming equal masses for the $\rho$ and $\omega$, there is an exact cancellation.

One finds that individually the $D^{*+} D^0, D^{*0} D^+$ states have a weak and repulsive interaction due to $J/\psi$ exchange 
and hence they do not bind by themselves,
but the coupled channels have the virtue of making a bound state possible.
Indeed, if we take the isospin combinations (our isospin doublets are $(D^+,\, -D^0)$ and $(D^{*+},\, -D^{*0})$)
\begin{equation}\label{eq:Iso}
\begin{split}
  |D^* D, I=0\rangle &= -\frac{1}{\sqrt{2}} (D^{*+} D^0 - D^{*0} D^+), \\[0.15cm]
  |D^* D, I=1, I_3=0\rangle &= -\frac{1}{\sqrt{2}} (D^{*+} D^0 + D^{*0} D^+),
\end{split} 
\end{equation}
we find (the indices indicating the isospin)
\begin{equation}\label{eq:C00}
C_{00} = \frac{1}{M_{J/\psi}^2}-\frac{1}{m_\rho^2}; ~~~C_{11} = \frac{1}{M_{J/\psi}^2}+\frac{1}{m_\rho^2};
~~~ C_{01} = 0;
\end{equation}
which means that we find an attraction, and not weak, in $I=0$ and repulsion in $I=1$.
An approximate solution can be obtained using the single channel with $I=0$,
which implies using average masses for $D^*$'s and $D$'s,
but we wish to be accurate and will use the coupled channels method with the exact masses.
The term with $\frac{1}{M_{J/\psi}^2}$ in Eq.~\eqref{eq:Cij} is kept since it comes from the extended local hidden gauge formalism, but being just a $6$\% of the $\frac{1}{m_\rho^2}$ term it can be safely neglected with no appreciable change in the results.

We solve then the Bethe-Salpeter equation in coupled channels and have in matrix form
\begin{equation}\label{eq:BSeq}
T=[1-VG]^{-1} \, V,
\end{equation}
with $G={\rm diag}[G_1, G_2]$, where $G_i$ are the $D^*D$ loop functions, 
which we regularize using dimensional regularization (DR) as in Ref.~\cite{Gamermann},
\begin{eqnarray}\label{eq:Gdimreg}
 G^{DR}_l &=&\frac{1}{16 \pi^2 } \bigg[ \alpha_H + \log\frac{M_1^2}{\mu^2}+\frac{M_2^2-M_1^2+s}{2s}\log\frac{M_2^2}{M_1^2} + \bigg. \nonumber\\
           & &   \bigg. \frac{p}{\sqrt{s}}  \left( \log(s-M_2^2+M_1^2+2p \sqrt{s}) \right.  \bigg.   \nonumber\\
           & & \bigg.  \left.  - \, \log(-s+M_2^2-M_1^2+2p \sqrt{s})      \right.   \bigg.     \nonumber\\
           & & \bigg. \left.   +\, \log(s+M_2^2-M_1^2+2p \sqrt{s})       \right.   \bigg.     \nonumber\\
           & &  \bigg. \left.  - \, \log(-s-M_2^2+M_1^2+2p \sqrt{s})  \right)  \bigg]
\end{eqnarray}
where $M_{1,2}$ are the masses of the two particles and $p$ the on shell three momentum of the two mesons, with the value of the renormalization energy scale $\mu=1500 \mev$ and the subtraction constant $\alpha_H$ having a value close to $\alpha_H=-1.15$.

Alternatively, one can also employ the cutoff regularization scheme
\begin{equation}\label{eq:Gcut}
G^{cut}_l = \int_{0}^{q_{max}} \frac{q^2 dq}{(2\pi)^2} \frac{\omega_1 + \omega_2}{\omega_1\omega_2 \left[(P^0)^2-(\omega_1 + \omega_2)^2+i\epsilon\right]} ,
\end{equation}
where $q_{max}$ stands for the cutoff in the three momentum, $\omega_i = \sqrt{{\vec{q}}^2 + M_i^2}$ and  ${P^0}^2=s$.\\

The value of $\alpha_H$ is fine tuned to get the experimental binding of $T_{cc}$ state.
Yet, to get a finite width for the state below the $D^{*} D$ thresholds we need to consider the width of the $D^*$ states.
This is accomplished performing a convolution of the $G$ functions with the spectral function (mass distribution) of the $D^*$ states,
as done in Ref.~\cite{Guanying} (see Eqs.(4),(5) of that reference),
with the width of the $D^*$ states showing the energy dependence:
\begin{eqnarray}\label{eq:Dstar1}
\Gamma_{D^{*+}}(M_{\rm inv})&=&\Gamma (D^{*+}) \left( \frac{m_{D^{*+}}}{M_{\rm inv}}\right)^2 \cdot    \nonumber \\
 & & \left[ \frac{2}{3}\left(\frac{p_\pi}{p_{\pi, \rm on}}\right)^3+\frac{1}{3}\left(\frac{p'_\pi}{p'_{\pi, \rm on}}\right)^3\right],
\end{eqnarray}
where $p_\pi$ is the $\pi^+$ momentum in $D^{*+} \to D^0 \pi^+$ decay with $D^{*+}$ mass $M_{\rm inv}$,
and $p_{\pi, \rm on}$ the same one with the physical mass of $D^{*+}$ taken from the PDG \cite{PDG}.
Analogously, $p'_{\pi}, p'_{\pi, \rm on}$ are the same magnitudes for $D^{*+} \to D^+ \pi^0$.
The width $\Gamma(D^{*+})$ is taken from the PDG, $\Gamma(D^{*+})=83.4 \; {\rm keV}$. For the $D^{*0}$, we take 
\begin{eqnarray}\label{eq:Dstar0}
\Gamma_{D^{*0}}(M_{\rm inv})&=&\Gamma (D^{*0})\left( \frac{m_{D^{*0}}}{M_{\rm inv}}\right)^2 \cdot    \nonumber \\
& & \left[0.647\left(\frac{p_\pi}{p_{\pi, \rm on}}\right)^3+0.353 \right],
\end{eqnarray}
where the second term corresponds to the $D^{*0} \to D^0 \gamma$ decay, 
which does not change appreciably with the small changes in $M_{\rm inv}$ of our problem,
and we have taken the branching fractions from the PDG and the value of $\Gamma (D^{*0})= 55.3 \;\textrm{keV}$ from Ref.~\cite{kungds}.
The values of $p_\pi, p_{\pi,\, {\rm on}}$ correspond now to the $D^{*0} \to D^0 \pi^0$ decay. The results are practically indistinguishable if we use $\Gamma (D^{*0})= 55.9 \;\textrm{keV}$ from Ref.~\cite{rosner} or $77.7\;\textrm{keV}$ from Ref.~\citep{Wangzhu}, indicating that the $D^{*+} D^0$ channel is the one playing a major role in the state given its proximity to the $D^{*+} D^0$ threshold. Indeed, if we change $\Gamma (D^{*0})$ from $55.9 \;\textrm{keV}$ to $77.7 \;\textrm{keV}$, the results change in the fourth decimal.\\

\begin{figure}[h]
\centering
\includegraphics[width=0.5\textwidth]{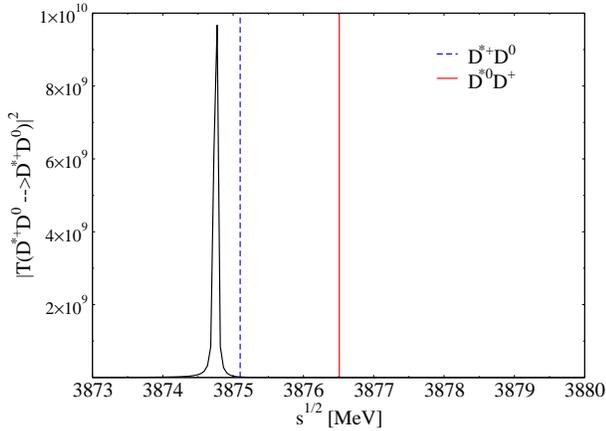}
\vspace{-0.2cm}
\caption{\small{$|T_{D^{*+} D^0,\, D^{*+}D^0}|^2$ as a function of $\sqrt{s}$. 
Dashed vertical line, $D^{*+}D^0$ threshold. Continuous vertical line, $D^{*0}D^+$ threshold.}}
\label{fig:T2}
\end{figure}

In Fig.~\ref{fig:T2} we show the results for $|T_{D^{*+} D^0,\, D^{*+}D^0}|^2$ as a function of $\sqrt{s}$ for the case of two channels using the mass of Eq.~\eqref{eq:mexp2} as input \footnote{The first version of the paper was done before Refs.~\cite{LHCb:2021vvq,LHCb:2021auc} were made public. We show these results first and later we show the new results if we take the value of Eq.~\eqref{eq:mpole} for $\delta m_{exp}$.}.

We have taken two subtraction constants $\alpha_H =-0.863$ for $D^{*+} D^0$ and $\alpha_H =-1.03$ for $D^{*0} D^+$,
and we find a neat peak around the experimental mass.
We see that the explicit consideration of the $D^*$ widths provides a width at the peak.
The width of the peak for the two channels case is 
\begin{equation}\label{eq:Gam3}
\Gamma\simeq 80 ~\textrm{keV}.
\end{equation}
This value is small compared with the experimental width of Eq.~\eqref{eq:mexp3}, even considering the large errors, and is about 60\% larger than those obtained in Refs.~\cite{shilingam} and \cite{xiegeng} of the order of $50~{\rm keV}$, using the couplings of $T_{cc}$ to the $D^* D$ components and the input of the $T_{cc}$ mass given by Eq.~\eqref{eq:mexp2}. The explicit consideration of the coupled channels with the convolutions done using the energy dependent widths is responsible for this increased width. However, it is about double than the value obtained in Ref.~\cite{LHCb:2021auc},  Eq.~\eqref{eq:wpole}, after the consideration of the experimental resolution. Nonetheless, we shall see later that when we use the mass of Eq.~\eqref{eq:mpole} as input, the width is considerably reduced and is in agreement with Eq.~\eqref{eq:wpole}.

It is interesting to see which are the couplings of the resonance in the case of two channels for the state corresponding to the peak of Fig.~\ref{fig:T2}. They are obtained from $T_{11}$ and $T_{12}$ as $g_{1}^2=\lim_{s \to s_R} (s-s_R)T_{11}$, $g_{2}=g_{1}T_{21}/T_{11}$.
In the easy case of neglecting the width of the $D^*$ states where the state appears as bound 
and there is no problem in defining the Riemann sheet we get
\begin{eqnarray}\label{eq:gi}
g_{T_{cc}, D^{*+}D^0}&=&3658.30~{\rm MeV} ; \nonumber \\
g_{T_{cc}, D^{*0}D^+}&=&-3921.04~{\rm MeV},
\end{eqnarray}
and, as we can see, they are basically opposite to each other indicating that we have indeed a quite good $I=0$ state, 
in spite of using different masses for the components and being close to thresholds. This finding is in agreement with the conclusions in Ref.~\cite{LHCb:2021auc}.

\begin{figure}[h]
\centering
\includegraphics[width=0.45\textwidth]{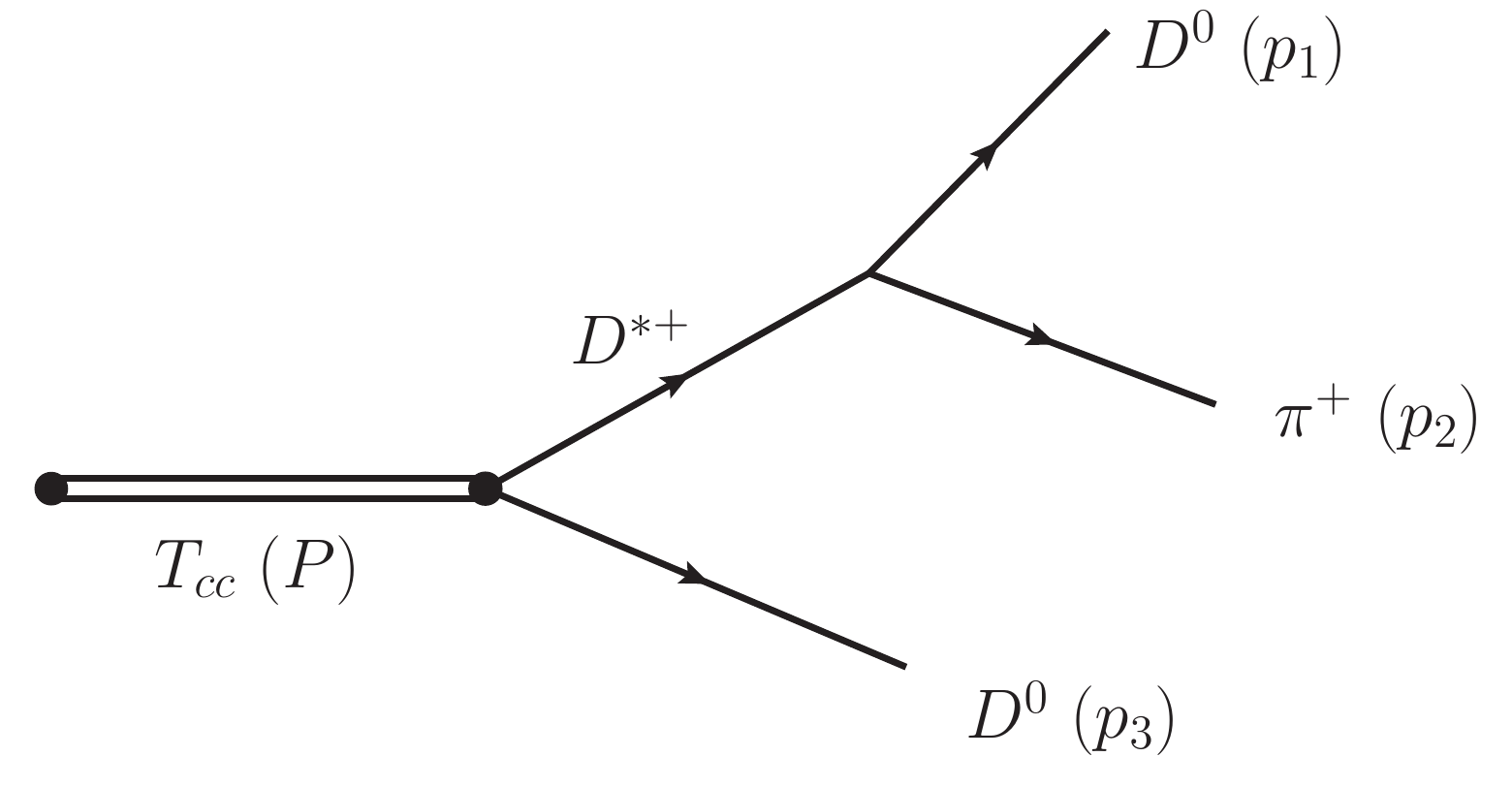}
\vspace{-0.2cm}
\caption{\small{Mechanism for $D^0\pi^+ D^0$ decay of the $T_{cc}$ state. The diagram with $D^{*0} D^+$ decay does not lead to final $D^0 D^0 \pi^+$.}}
\label{fig:decay}
\end{figure}

Next we make a study of the $D^0 D^0 \pi^+$ mass distribution in the decay of the resonance,
the channel observed in the experiment.
This corresponds to a diagram like the one depicted in Fig.~\ref{fig:decay}.  

\begin{figure}[h]
\centering
\includegraphics[width=0.5\textwidth]{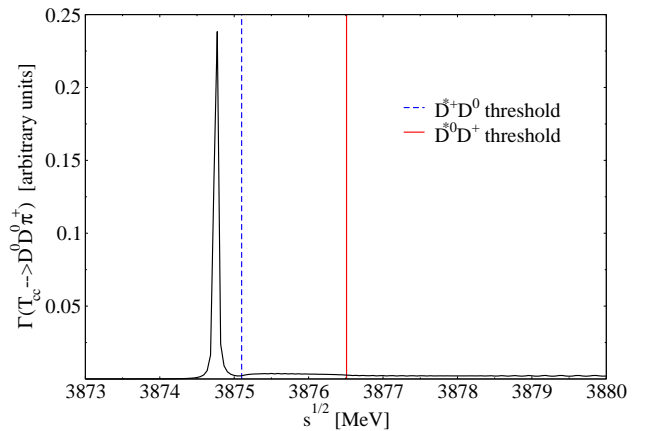}
\vspace{-0.2cm}
\caption{\small{$\Gamma(\sqrt{s})$ for the decay of the $T_{cc}$ into $D^0 D^0 \pi^+$.}}
\label{fig:Gamm}
\end{figure}

The decay of whichever object producing the $T_{cc}$ and decaying to $D^0 D^0 \pi^+$ can be obtained with the standard formula
\begin{equation}\label{eq:dGam}
  \dfrac{{\textrm d} \Gamma}{{\textrm d}M_{12}^2 \; {\textrm d}M_{23}^2}=\frac{1}{2} \, \dfrac{1}{(2\pi)^3} \, \dfrac{1}{s^{3/2}}\, |t|^2,
\end{equation}
where $t$ is obtained from the diagram of Fig.~\ref{fig:decay},
symmetrizing over the two $D^0$ momenta and the factor $\frac{1}{2}$ is added in the formula.
The amplitude $t$ for this process is given by
\begin{eqnarray}\label{eq:t}
  t &=& {\cal{C}}T_{D^{*+}D^0,D^{*+}D^0}(\sqrt{s}) \left[  \frac{\vec \epsilon \cdot (\vec p_1 -\vec p_2) }{M^2_{12}-m^2_{D^{*+}}+i M_{12} \Gamma_{D^{*+}}(M_{12})} \right.  \nonumber\\
   && + \left. \frac{\vec \epsilon \cdot (\vec p_3 -\vec p_2)}{M^2_{23}-m^2_{D^{*+}}+i M_{23}\, \Gamma_{D^{*+}}(M_{23})} \right],
\end{eqnarray}
where $\cal{C}$ is an arbitrary constant and $\vec \epsilon$ stands for the polarization vector of the $T_{cc} (1^+)$.
Upon summing over the polarizations $\vec \epsilon$ in $|t|^2$ for $T_{cc}$ at rest, we have terms
\begin{equation}\label{eq:pol}
\sum_{\textrm{pol.}} (\vec \epsilon \cdot \vec q_1)\, (\vec \epsilon \cdot \vec q_2) =  \epsilon_\mu \,q_1^\mu \, \epsilon_\nu \,q_2^\nu
=\left( -g^{\mu\nu} + \dfrac{P_\mu \, P_\nu}{M^2_{T_{cc}}}  \right)\, q_1^\mu \, q_2^\nu,
\end{equation}
where, $\vec q_1$ and $\vec q_2$ can be $(\vec p_1 -\vec p_2)$ and $(\vec p_3 -\vec p_2)$.
We convert the terms $\vec q_1 \cdot \vec q_2$ into invariants which can be written in terms of $M_{12}$ and $M_{23}$ 
using that $M^2_{12} + M^2_{12} +M^2_{23}=M^2_{T_{cc}}+ m^2_{D^0}+ m^2_{D^0}+m^2_{\pi^+}$.
Integrating Eq.~\eqref{eq:dGam} over $M_{12}$ and $M_{23}$, using the limits of the PDG for the Dalitz boundary,
we obtain the mass distribution $\Gamma(\sqrt{s})$ shown in Fig.~\ref{fig:Gamm}.\\

\begin{figure}[h]
\centering
\includegraphics[width=0.5\textwidth]{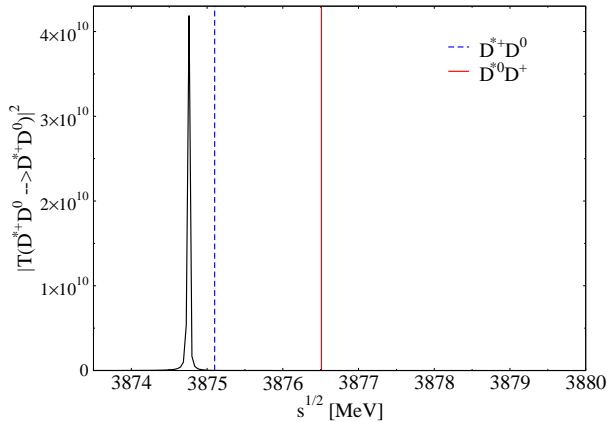}
\vspace{-0.2cm}
\caption{\small{$|T_{D^{*+} D^0,\, D^{*+}D^0}|^2$ as a function of $\sqrt{s}$. 
Dashed vertical line, $D^{*+}D^0$ threshold. Continuous vertical line, $D^{*0}D^+$ threshold.}}
\label{fig:T2_new}
\end{figure}

\begin{figure}[h]
\centering
\includegraphics[width=0.5\textwidth]{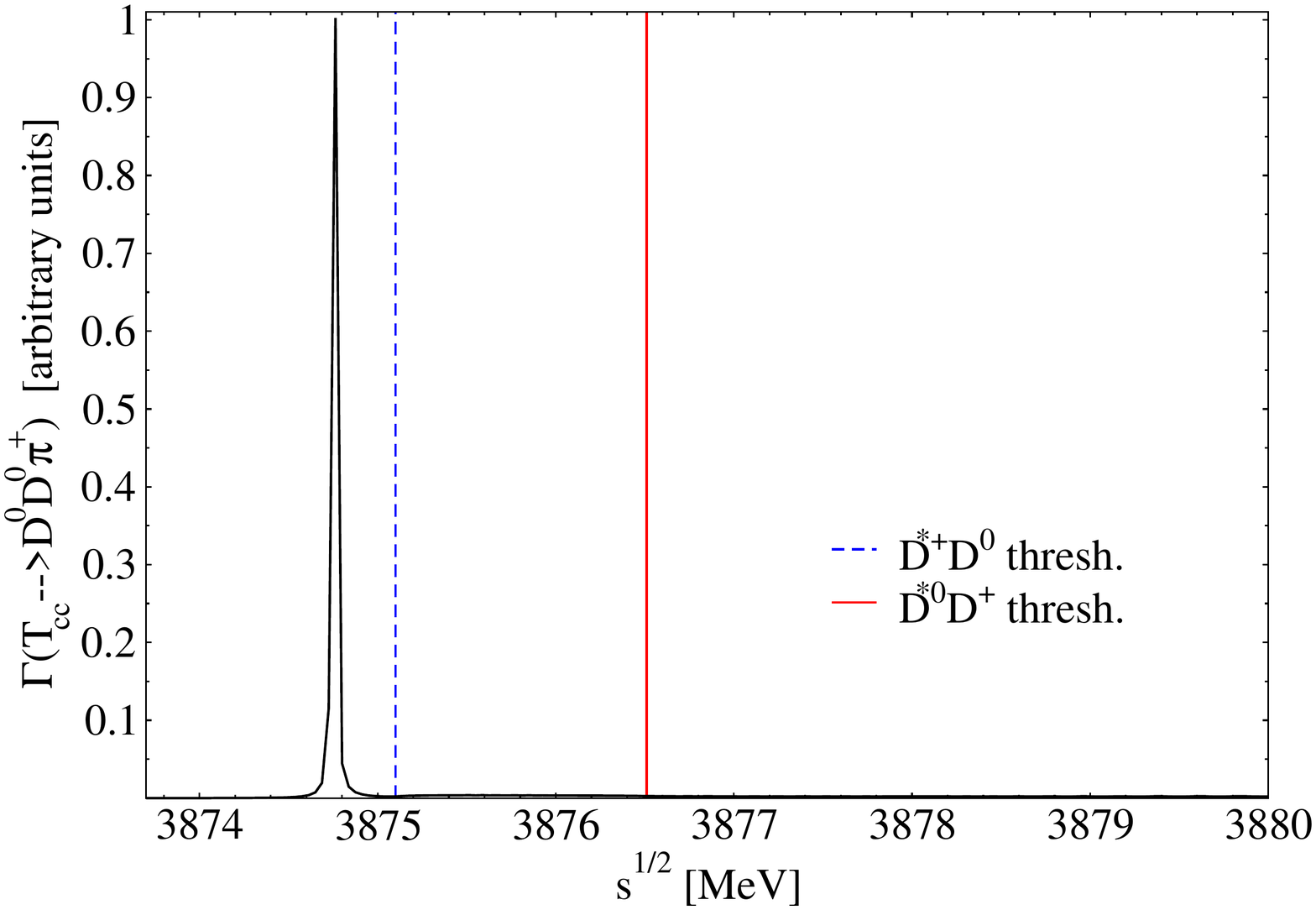}
\setlength{\unitlength}{0.0275\linewidth}
    \begin{picture}(100,0)    
       \put(17.4,13.3){\includegraphics[width=16\unitlength]{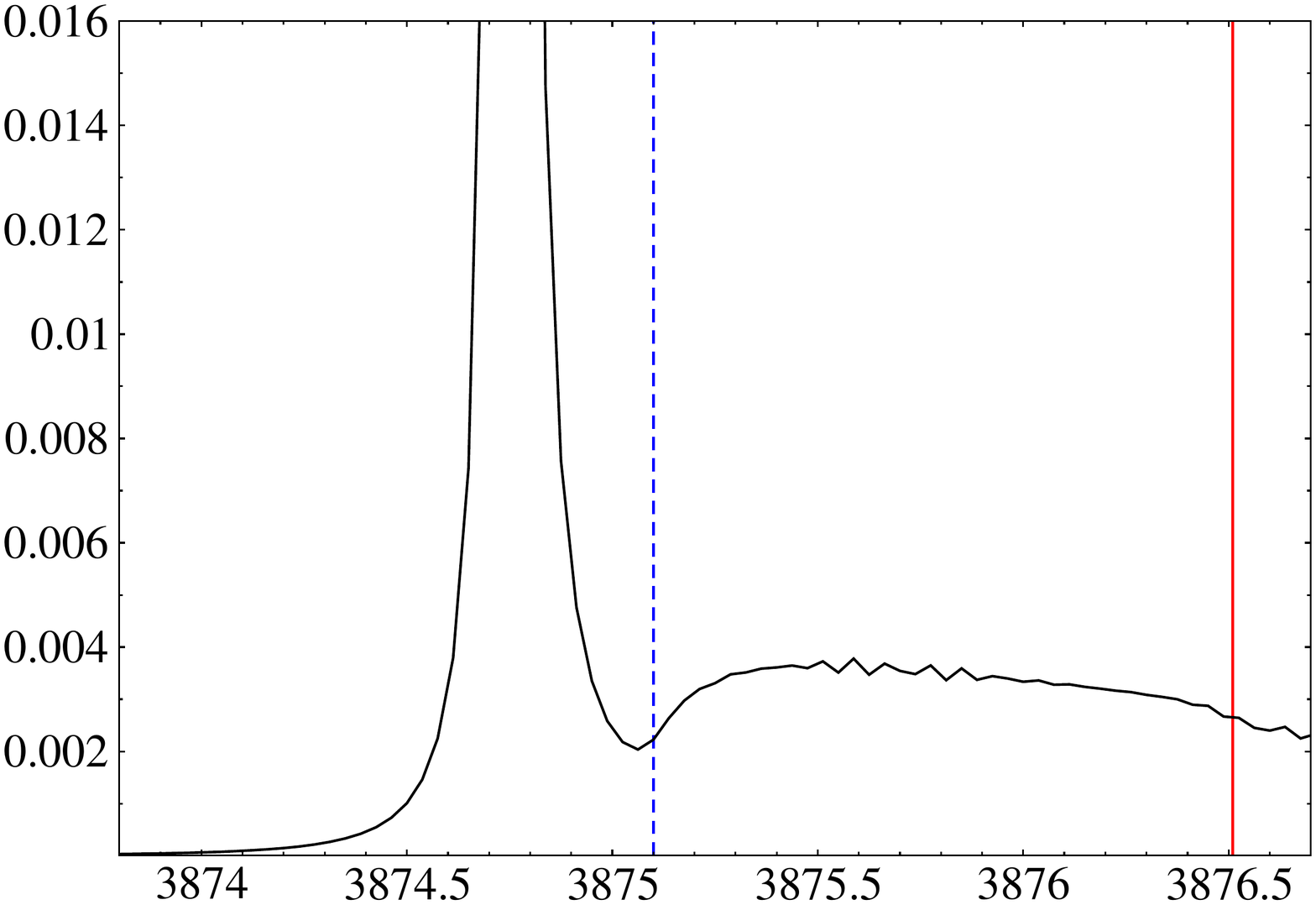}}
    \end{picture}
\caption{\small{$\Gamma(\sqrt{s})$ for the decay of the $T_{cc}$ into $D^0 D^0 \pi^+$.The inset in the figure is a zoom to illustrate the mass distribution between $D^{*+} D^0$ and $D^{*0} D^+$ thresholds.}}
\label{fig:Gamm_new}
\end{figure}

Next we repeat the calculations using the mass of Ref.~\cite{LHCb:2021auc}, Eq.~\eqref{eq:mpole}, as input. The new mass is obtained taking $\alpha_H =-0.870$ for $D^{*+} D^0$ and $\alpha_H =-1.03$ for $D^{*0} D^+$. In Fig.~\ref{fig:T2_new} and \ref{fig:Gamm_new}, we show now $|T_{D^{*+} D^0,\, D^{*+}D^0}|^2$ and the $D^0 D^0 \pi^+$ spectrum respectively. We see that now the width at half the strength of the peak is   
\begin{equation}\label{eq:Gam_new}
\Gamma\simeq 43 ~\textrm{keV}, 
\end{equation}
which is in good agreement with the findings of Ref.~\cite{LHCb:2021auc} in Eq.~\eqref{eq:wpole}. The shape of the spectrum in Fig.~\ref{fig:Gamm_new} is practically identical to the one obtained from analysis of the LHCb data in Ref.~\cite{LHCb:2021auc} taking into account the experimental resolution, and also has a small contribution between the two thresholds, which is tied to the $D^0 D^0 \pi^+$ spectrum and does not show up in $|T|^2$.

The couplings that we get now are 
\begin{eqnarray}
g_{T_{cc}, D^{*+}D^0}&=&3884.68~{\rm MeV} ; \nonumber \\
g_{T_{cc}, D^{*0}D^+}&=&-4144.26~{\rm MeV} .
\end{eqnarray}\label{eq:gi_new}

One should stress that these couplings are also consistent with those determined by the experimental analysis \cite{LHCb:2021auc}
\begin{equation*}
 |g| >\frac{5.1}{\sqrt{2}}=3.6~\textrm{GeV} \; \left(\frac{4.3}{\sqrt{2}}=3.0~\textrm{GeV} \right) ~~\textrm{at}~~90(95\%)~\textrm{CL}
\end{equation*}
and close by to them.\\

It is also interesting to see what do we get if we assume just one channel with exact $I=0$ as assumed in the analysis of Ref.~\cite{LHCb:2021auc}, Eq.~\eqref{eq:Iso}, and averaged masses for the $D^*$ mesons and $D$ mesons. The calculations are done with a single channel with $V$ given by Eq.~\eqref{eq:Vij} with $C_{00}$ of Eq.~\eqref{eq:C00}. The convolved $G$ function is taken as the average of $G_{D^{*+}D^0}$ and $G_{D^{*0}D^+}$ using average masses and the values of the widths, Eqs.~\eqref{eq:Dstar1},~\eqref{eq:Dstar0} for each $G$ function and a single $\alpha_H$ for the two channels. The results do not change qualitatively from those of  Figs.~\ref{fig:T2_new} and \ref{fig:Gamm_new}, only the width of the state is now $\Gamma\simeq 55 ~\textrm{keV}$ and the strength of the $D^0 D^0 \pi^+$ spectrum between the thresholds, while still very small is about three times bigger than before, as a consequence of the change of the thresholds when employing average masses.\\

It is interesting to remark the coincidence in practice of the approach of Ref.~\cite{LHCb:2021auc} making fits to the data with a unitary Breit-Wigner amplitude and our approach based upon the use of the $D^*D$ scattering matrix, with the $G$ function convolved to account for the $D^*$ widths, and the explicit mechanism of Fig.~\ref{fig:decay} to account for the $D^0 D^0 \pi^+$ production. The formalisms look rather different but the physics contained in them coincide. We would expect the same results for other mass distributions discussed in Ref.~\cite{LHCb:2021auc}. In particular, the most relevant, the lack of any signal in channels related to a possible $I=1$ state, is guaranteed in our approach since we already showed that we do not get any state for $I=1$, since the interaction is repulsive there.\\

Another interesting information can be extracted using the cutoff regularization of Eq.~\eqref{eq:Gcut} and we obtain $q_{max}=415$~MeV when taking the mass of Eq.~\eqref{eq:mexp2} and $418.6$~MeV when using the mass of Eq.~\eqref{eq:mpole}. This value is somewhat small compared with the $600$~MeV that one needs in the study of the low lying scalar mesons ($\sigma$, $f_0$, $a_0$) of $q_{max}=600$~MeV \cite{Liang:2014tia}. Since, in one channel one has $T=\left[V^{-1}-G\right]^{-1}$, a decrease of $|V|$ by means of a repulsive interaction induces an increase in $|G|$ to get the pole at the same place, implying a larger cut off. We perform the test of increasing the repulsion of $J/\psi$ exchange by a factor $3$ in Eq.~\eqref{eq:Cij}, which still makes this contribution only about $18$\% of the $\rho$ exchange and we need $q_{max}=476$~MeV, closer to the $600$~MeV used in Ref.~\cite{Liang:2014tia}. Note that our approach generates an accurate interaction from the exchange of light vectors, which is consistent with heavy quark symmetry, but subleading terms, like the exchange of heavy vectors which do not follow that rule, are less accurate.

\section{Summary}

In summary, we get a molecular state of $D^{*+} D^0, D^{*0} D^+$, 
with a mixing that corresponds very approximately to an $I=0$ state. 
No signal is seen for the orthogonal, approximately $I=1$ state, as one can see in Fig. 2. 
This is in contrast to the suggestion made in Ref.~\cite{shilinewnew} that two states could come from isospin mixing. 
Actually, the small bump in the $D^0 D^0 \pi^+$ spectrum suggested in Ref.~\cite{shilinewnew} as a possible new state, 
also shows up in our spectrum of $D^0 D^0 \pi^+$ in Fig.~\ref{fig:Gamm},  
but since this small bump does not appear in $|T|^2$ in Fig.~\ref{fig:T2}, 
it has to be associated to the decay channel $D^0 D^0 \pi^+$ and its phase space 
and not to a physical state. 
This also means that measuring other decay channels would provide new and valuable information, as it has been shown in Ref.~\cite{LHCb:2021auc}. \\

The analysis of the LHCb data done in Ref.~\cite{LHCb:2021auc}, considering the experimental resolution and using a unitary Breit Wigner amplitude, has been most useful since it allows to  compare with theoretical calculations as the one we have done. We have shown that when using the mass obtained in Ref.~\cite{LHCb:2021auc} as input, the width obtained for the $T_{cc}$ state and the $D^0 D^0 \pi^+$ mass distribution are in remarkable agreement with the results reported in  Ref.~\cite{LHCb:2021auc}.

\section{ACKNOWLEDGEMENT}
This work is partly supported by the National Natural Science Foundation of China under Grants No. 11975083 and No. 12047567.
This work is also partly supported by the Spanish Ministerio de Economia y Competitividad (MINECO)
and European FEDER funds under Contracts No. FIS2017-84038-C2-1-P B, PID2020-112777GB-I00,
and by Generalitat Valenciana under contract PROMETEO/2020/023.
This project has received funding from the European Union Horizon 2020 research and innovation programme
under the program H2020-INFRAIA-2018-1, grant agreement No. 824093 of the ``STRONG-2020" project.
The present work has been also partially supported by the Czech Science Foundation, GACR Grant No. 19-19640S.


\end{document}